\documentclass[conference,twoside,letterpaper]{IEEEtran}
\IEEEoverridecommandlockouts

\ifCLASSINFOpdf
\else
\fi

\usepackage[cmex10]{amsmath}
\usepackage{cite}
\usepackage{mathtools}
\usepackage{flushend}
\usepackage{url}
\usepackage{enumerate}
\usepackage{comment}
\usepackage{tabularx}
\usepackage{color}
\begin{document}

% vspacet -9 ja -5 isommilla marginaaleilla
\title{Reference Receiver Based Digital Self-Interference Cancellation in MIMO Full-Duplex Transceivers\vspace{-8mm}}%\vspace{-9mm}

\author{\IEEEauthorblockN{Dani~Korpi,
Lauri~Anttila,
and~Mikko~Valkama\vspace{-3mm}}
\\
\IEEEauthorblockA{Department of Electronics and Communications Engineering, Tampere University of Technology, Finland\\ e-mail: dani.korpi@tut.fi, lauri.anttila@tut.fi, mikko.e.valkama@tut.fi\vspace{-4mm}}%\vspace{-5mm}
\thanks{The research work leading to these results was funded by the Academy of Finland (under the project \#259915 "In-band Full-Duplex MIMO Transmission: A Breakthrough to High-Speed Low-Latency Mobile Networks"), the Finnish Funding Agency for Technology and Innovation (Tekes, under the project "Full-Duplex Cognitive Radio"), the Linz Center of Mechatronics (LCM) in the framework of the Austrian COMET-K2 programme, and Emil Aaltonen Foundation. The research was also supported by the Internet of Things program of DIGILE (Finnish Strategic Centre for Science, Technology and Innovation in the field of ICT), funded by Tekes.}}% 

\maketitle

\begin{abstract}

In this paper we propose and analyze a novel self-interference cancellation structure for in-band MIMO full-duplex transceivers. The proposed structure utilizes reference receiver chains to obtain reference signals for digital self-interference cancellation, which means that all the transmitter-induced nonidealities will be included in the digital cancellation signal. To the best of our knowledge, this type of a structure has not been discussed before in the context of full-duplex transceivers. First, we will analyze the overall achievable performance of the proposed cancellation scheme, while also providing some insight into the possible bottlenecks. We also provide a detailed formulation of the actual cancellation procedure, and perform an analysis into the effect of the received signal of interest on self-interference coupling channel estimation. The achieved performance of the proposed reference receiver based digital cancellation procedure is then assessed and verified with full waveform simulations. The analysis and waveform simulation results show that under practical transmitter RF/analog impairment levels, the proposed reference receiver based cancellation architecture can provide substantially better self-interference suppression than any existing solution, despite deploying only low-complexity linear digital processing.

\end{abstract}
%\vspace{-2mm}

% \fi
\IEEEpeerreviewmaketitle

\section{Introduction}
%\vspace{-1mm}

Full-duplex radio communications with simultaneous transmission and reception at the same radio frequency (RF) carrier has recently gained considerable interest among researchers. It has the potential to significantly improve the efficiency and flexibility of RF spectrum usage, which makes it an appealing concept when trying to increase the data rates of the current systems while retaining the utilized resources. There have already been several promising demonstration-type implementations of such full-duplex radio transceivers \cite{Knox12,Bharadia13,Duarte12,Jain11}. In addition to practical demonstrations, there has also been a large number of theoretical studies investigating the boundaries of in-band full-duplex communications under various circuit impairments and deployment scenarios \cite{Korpi133,Korpi13,Syrjala13,Riihonen122}.

%It has been shown, firstly, that phase noise can form a serious bottleneck in full-duplex transceivers \cite{Sahai12,Syrjala13}. Especially when assuming separate local oscillator (LO) signals for up- and downconversion, the level of phase noise can dramatically degrade the final signal-to-noise-plus-interference ratio (SINR). However, under the assumption that all the mixers are fed with the same LO signal, the phase noise is not likely to be the most crucial issue in the context of full-duplex transceivers.

Among the various circuit imperfections, nonlinear distortion is one central impairment which has been observed to be a significant issue in in-band full-duplex transceivers \cite{Korpi13,Bharadia13,Ahmed13,Anttila13}. Especially the transmitter (TX)-induced nonlinearities can heavily limit the amount of achievable self-interference (SI) suppression if only linear processing is utilized in the cancellation stages \cite{Anttila13,Korpi13}. For this reason, several studies have recently proposed nonlinear SI cancellation techniques, which would also be able to attenuate nonlinearly distorted SI signals \cite{Anttila13,Bharadia13,Ahmed13}. These methods rely on nonlinear digital signal processing algorithms which model and attenuate the nonlinearities after the analog-to-digital conversion. Thus, with the cost of increased computational and digital hardware complexity, it is already possible to mitigate the problem of nonlinear distortion in full-duplex transceivers.

In addition to nonlinear distortion, it has also been observed recently that the effect of IQ imbalance can be very harmful in in-band full-duplex communications, when assuming a practical image rejection ratio for the IQ mixers \cite{Korpi133}. However, by utilizing widely-linear processing in the digital domain, it is possible to suppress also the conjugate SI caused by the IQ imbalance. Thus, similar to nonlinear distortion, it is possible to mitigate the problems caused by IQ imbalance by increasing the computational complexity of digital SI cancellation \cite{Korpi133}.

In this paper, we take an alternative path and propose \emph{a novel architecture for SI cancellation, which allows for efficient and accurate digital cancellation in the presence of most prominent TX chain nonidealities, while keeping the computational complexity of the digital calculations at the level of typical linear processing.} For generality, we focus on a MIMO full-duplex transceiver scenario and develop a cancellation architecture where additional reference receiver branches, together with linear digital processing, are deployed to create the digital baseband cancellation signals. This way, high accuracy in digital SI cancellation can be achieved, since also the nonlinear distortion and IQ images of the TX chains are suppressed, without increasing the digital processing complexity. Especially in MIMO devices, avoiding the need for complicated nonlinear spatio-temporal type of digital processing in SI cancellation is a substantial benefit. Also, to the best of our knowledge, this is the first SI cancellation approach capable of modeling impairment coexistence in in-band full-duplex transceivers. Namely, as our results show, it is able to attenuate both the TX-induced SI mirror image and PA-induced nonlinear distortion in the digital domain, thereby outperforming all the existing digital cancellation algorithms when assuming realistic TX chains.

In summary, the novelty and contributions of this work are:
\begin{itemize}
\item Propose and formulate the reference receiver based full-duplex MIMO transceiver concept
\item Analyze the SI cancellation performance of the proposed structure under practical RF circuit imperfections
\item Derive the Cram\'{e}r--Rao Lower Bound on coupling channel parameter estimator variance
\item Establish analytically and through simulations that the proposed structure outperforms existing linear and nonlinear SI cancellation solutions under practical RF imperfections
\end{itemize}

The rest of this paper is organized as follows. In Section~\ref{sec:architectures}, some of the previous full-duplex transceiver structures are briefly reviewed. The proposed structure utilizing reference receiver aided digital cancellation is then presented in Section~\ref{sec:proposed_model}, together with elementary transceiver system calculations and estimation theoretic results regarding its performance. After this, in Section~\ref{sec:simulations}, the achievable performance of the proposed structure is further analyzed and verified with full waveform simulations. Finally, conclusions are drawn in Section~\ref{sec:conc}.

%\vspace{-1.1mm}
\section{Previous Full-duplex Transceiver Architectures}
\label{sec:architectures}
%\vspace{-0.4mm}

There have already been several functional implementations of full-duplex transceivers, and the overall structure of all the implemented transceivers has been relatively similar. The transmitter may have separate transmit and receive antennas \cite{Jain11,Duarte12} or only one antenna for both transmission and reception \cite{Knox12,Cox13}. In the receiver, the SI signal is first actively attenuated in the RF domain by subtracting the known transmit signal from the received signal. There have been two alternative structures for this RF cancellation stage. In \cite{Jain11,Bharadia13}, the cancellation is done by subtracting a properly delayed, attenuated and phase-rotated, or more generally, a filtered version of the transmitter output signal from the received signal at the input of the receiver (RX) chain. In \cite{Duarte12}, on the other hand, the RF cancellation signal is generated from the digital transmit samples and then fed through an additional reference transmitter chain, after which it is subtracted from the received signal at the input of the RX chain. The former structure has the benefit of not requiring an additional transmitter chain, but the latter method is able to do the channel estimation and processing in the digital domain, which decreases the complexity of the RF circuitry.

Usually, RF cancellation is the only form of active SI attenuation performed in the analog domain. However, recently there have been also some experiments regarding another analog SI cancellation stage before the analog-to-digital converter (ADC) \cite{Kaufman132,Hua13}. The idea behind these techniques is to inject a cancellation signal to the input of the ADC, and thereby attenuate the level of SI before quantization. This means that there will be more bits left for the actual signal of interest (SOI) since the SI signal will not reserve as large a portion of the available dynamic range. Thus, an important benefit of analog baseband cancellation is the decreased requirements for the ADC, but it comes with the cost of the additional cancellation circuitry. The choice of the reference signal for the analog baseband cancellation plays also a significant role. In \cite{Kaufman132}, the reference signal is taken directly from the output of the TX digital-to-analog converter (DAC), whereas in \cite{Hua13} it is taken from the RF output of the transmitter. The latter technique has the benefit of also attenuating the TX-induced nonidealities in the baseband cancellation stage.

However, if it is assumed that the first analog cancellation stage is able to attenuate the SI to a tolerable level in terms of the ADC dynamic range, it could be possible to utilize the structure presented in \cite{Hua13}, but only with digital SI cancellation, instead of analog baseband cancellation. As already mentioned, in this case the reference signals for digital cancellation would include all the possible TX-induced nonidealities, and thereby they would be attenuated by digital cancellation, even if only linear processing was performed. In this paper, we will formulate and analyze this type of a novel full-duplex transceiver structure in the general MIMO transceiver context.

\section{Proposed MIMO Full-duplex Transceiver with Reference Receivers}
\label{sec:proposed_model}

\begin{figure*}[!t]
\centering
\includegraphics[width=0.8\textwidth]{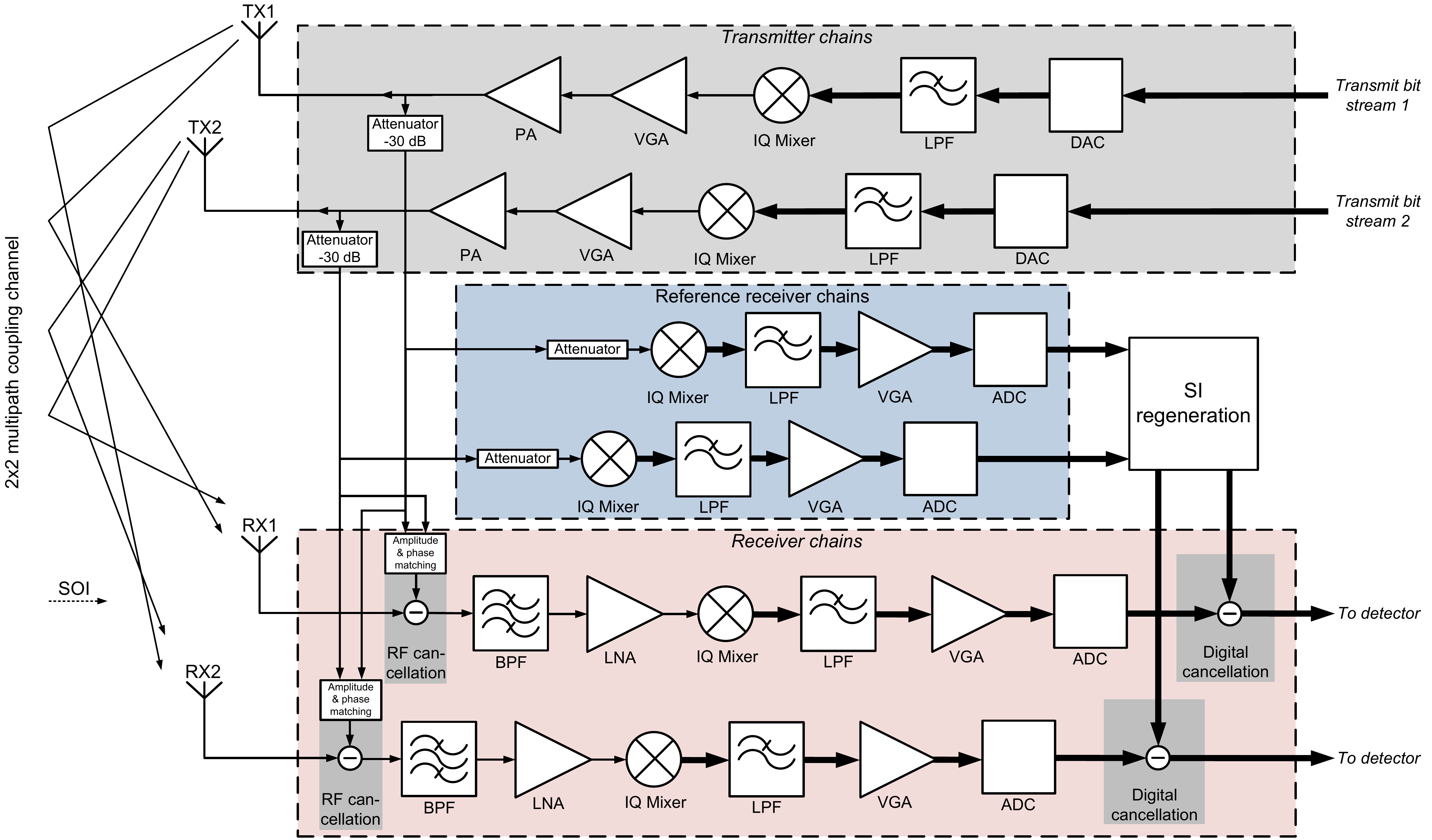}
\caption{A block diagram of the considered MIMO full-duplex transceiver, which employs reference receiver chains to generate a digital cancellation signal.}
\label{fig:block_diagram}
\vspace{-4mm}
\end{figure*}

A fairly similar MIMO full-duplex transceiver structure to the one presented in \cite{Korpi132} is proposed in this study. The main difference between the structures is in the practical implementation of digital SI cancellation. More specifically, in this paper we present and analyze a novel approach that utilizes reference RX chains, whose purpose is to provide the reference signals for digital SI cancellation such that TX-induced nonidealities can be efficiently suppressed with linear digital processing. The structure of the proposed 2x2 MIMO full-duplex transceiver is presented in Fig.~\ref{fig:block_diagram}.

As already mentioned, the most crucial benefit of this cancellation approach is that all the possible sources of TX chains' distortion are included in the digital cancellation signal, unlike in the traditional methods where the cancellation signal is constructed from the original transmitted data samples \cite{Jain11,Korpi133}. However, this benefit comes at a cost of $N_{tx}$ additional RX chains, where $N_{tx}$ is the number of transmit antennas. {\color{black}Therefore, the proposed structure essentially decreases the level of computational complexity in the digital domain by utilizing additional hardware in the analog domain.} Note that it is not necessary to have an LNA in these reference RX chains, as their input signals are already sufficiently powerful. A variable gain amplifier (VGA) is still required to be able to cope with different transmit power levels. %  {\color{red}In fact, having the LNA in the reference RX chains would most likely result in a very noisy reference signal, as the reference RX chain input signals would have to be heavily attenuated to avoid saturating the LNA.} 

\subsection{Quantifying the Analog Performance}

One aspect having an impact on the performance of the proposed structure are the possible nonidealities occurring in the reference RX chains. These nonidealities may distort the digital cancellation signal, and thus reduce the accuracy of SI regeneration. To analyze the effect of the different impairments on the performance of the full-duplex transceiver, let us model these nonidealities as additive noise sources. The approximate power of each distortion component can be determined with basic system calculations, as was demonstrated in \cite{Korpi13}. To make the analysis tractable, it is further assumed that all the distortion components are independent, which means that the total power of the signal consists of the sum of the powers of the different signal components. This condition might not always be met in reality, but it is a feasible approximation for the purpose of this analysis.

\subsubsection{Power Levels of the Signal Components}
Now, using the same procedures as in \cite{Korpi13}, we arrive with the following equations for the approximate powers of the different signal components \emph{at the detector input of an individual RX chain}:
\begin{align}
&g_{rx} = \frac{p_{target}}{\frac{N_{tx} p_{tx}}{a_{ant}}\left(\frac{1}{a_{RF}}+\frac{p_{tx}^2}{a_{RF}\mathit{iip3}_{PA}^2 g_{PA}^2} \right)+p_{soi,in}} \label{eq:g_rx}\\[3mm]
&p_{soi} = g_{rx} p_{soi,in}\label{eq:p_soi}\\[3mm]
&p_{n} = F_{rx}g_{rx}p_{th} \label{eq:p_n}\\[3mm]
&p_{si} = \frac{N_{tx} g_{rx} p_{tx}}{a_{ant} a_{RF} a_{dig}} \phantom{||xxxxxxxxxxxxxxxxxxxxxxxxxxx}\label{eq:p_si}%\\[3mm]
\end{align}
\begin{align}
&p_{si,im} = \frac{g_{rx} p_{tx}}{a_{ant} a_{RF}}\left( \frac{N_{tx}}{a_{dig} \mathit{irr}_{tx}} +\frac{N_{tx}+1}{ \mathit{irr}_{rx}} \right) \label{eq:p_siim}\\[3mm]
&p_{nl,tx} = \frac{N_{tx} g_{rx} p_{tx}^3}{\mathit{iip3}_{PA}^2 g_{PA}^2 a_{ant} a_{RF} a_{dig}} \label{eq:p_nltx}\\[3mm]
&p_{nl,rx} = \frac{N_{tx} g_{rx} p_{tx}^2}{a_{ant}^2 a_{RF}^2} \nonumber\\
&\times \left[(N_{tx}+1) \left(\frac{g_{LNA}}{\mathit{iip2}_{mixer}}+\frac{g_{LNA}g_{mixer}}{2\mathit{iip2}_{VGA}} \right)\right.\nonumber\\
&\left. {} +\frac{(N_{tx}^2+1)p_{tx}}{a_{ant}a_{RF}} \left(\left(\frac{g_{LNA}}{\mathit{iip3}_{mixer}}\right)^2+\left(\frac{g_{LNA}g_{mixer}}{2\mathit{iip3}_{VGA}} \right)^2 \right)\right.\nonumber\\
&\left. {} + \frac{N_{tx}^2 p_{tx}}{a_{ant}a_{RF} \mathit{iip3}_{LNA}^2} \right] \label{eq:p_nlrx}\\[3mm]
&p_{q,tot} = p_{target} \left(\frac{1}{\mathit{snr}_{ADC}} + \frac{N_{tx}}{\mathit{snr}_{ADC,ref}} \right) \label{eq:p_qtot}\text{,}%\\[3mm]
\end{align}
where $g_{rx}$ is the total gain of the RX chain, $p_{soi}$ is the power of the signal of interest, $p_n$ is the total thermal noise power, $p_{si}$ is the power of the linear self-interference signal, $p_{si,im}$ is the power of the conjugate SI, caused by transmitter, receiver, and reference receiver IQ imbalance, $p_{nl,tx}$ is the power of the TX-induced nonlinear distortion, $p_{nl,rx}$ is the power of the nonlinear distortion produced in the actual and reference RX chains, and $p_{q,tot}$ is the total power of the quantization noise, caused by both the actual and reference RX chains. In addition, $p_{tx}$ is the transmit power of a single transmitter, $p_{target}$ is the target power of the total signal at the input of the RX chain ADC, $a_{ant}$ is the amount of antenna attenuation experienced by each transmit signal before reaching a receive antenna, $a_{RF}$ is the amount of RF cancellation, $a_{dig}$ is the amount of digital cancellation, $N_{tx}$ is the number of transmit antennas, $g_{k}$ is the gain of the component $k$ when $k = \{\mathit{LNA}\text{, }\mathit{mixer}\text{, }\mathit{VGA}\text{, }\mathit{PA} \}$, $\mathit{iip2}_{k}$ is the 2nd-order intercept point of the component $k$, $\mathit{iip3}_{k}$ is the 3rd-order intercept point of the component $k$, $p_{soi,in}$ is the power of the signal of interest at the input of the RX chain, $F_{rx}$ is the noise factor of the RX chain, $p_{th}$ is the thermal noise power at the input of the RX chain, $\mathit{irr}_{tx}$ is the image rejection ratio of the transmitter, $\mathit{irr}_{rx}$ is the image rejection ratio of the receiver, $\mathit{snr}_{ADC}$ is the dynamic range of the RX chain ADC, and $\mathit{snr}_{ADC,ref}$ is the dynamic range of the reference RX chain ADC. Note that in \eqref{eq:g_rx}--\eqref{eq:p_qtot} all the variables are assumed to be in linear power units.

The previous equations have been obtained under the assumption that the reference RX chains are otherwise similar to the real RX chains, except for the lack of the LNA, and the amount of bits at the ADC. This means that when the transmit signals are attenuated by $\frac{g_{LNA}}{a_{ant} a_{RF}}$, they are at the correct level at the input of each reference RX chain. Thus, the LNAs are replaced by attenuators, but otherwise similar components are used for the reference RX chains. The benefit of this approach is that it is not necessary to have a separate automatic gain control (AGC) algorithm for the reference RX chains, as their variable gain amplifiers (VGA) are able to use the same gain as the VGAs of the actual RX chains. This also allows for a significantly more illustrative analysis of the transceiver structure.

In addition to the above signal components, there are also additional sources of distortion but in this analysis we omit them as they are insignificant with respect to the more powerful components. One such distortion component is the phase noise, which has been shown to be a significant issue under certain circumstances \cite{Syrjala13,Sahai12}. However, in this analysis we assume that all the mixers are fed by the same oscillator signal, which means that the effect of the phase noise is insignificant \cite{Syrjala13}.

\subsubsection{Example Parameter Values}
Now, to determine the performance bounds of the proposed full-duplex transceiver structure, let us first define some feasible parameter values for the individual components. The values are shown in Tables~\ref{table:system_parameters} and~\ref{table:parameters}, and they are chosen to represent a realistic full-duplex transceiver based on earlier literature and specifications \cite{Parssinen99,Behzad07,Gu06,Jain11,Duarte12,LTE_specs}. In this analysis it is assumed that the RX chain has been calibrated properly to achieve a fairly high image rejection ratio (IRR). This is a feasible assumption for a typical direct-conversion transceiver \cite{Anttila08}.

In addition, in this section it is assumed that the accuracy of linear digital cancellation is perfect, i.e., $a_{dig} \to \infty$, to obtain proper insight into the performance boundaries of the proposed structure. In reality, the accuracy of digital cancellation is of course never perfect but, with reasonably good channel estimate quality, it is possible to attenuate all the signal components included in the reference signal well below the noise floor. Thus, in the context of determining the maximum achievable performance, perfect digital cancellation is a feasible assumption. Due to the novel technique for obtaining the reference signals, perfect digital cancellation now means that also the TX-induced nonlinear distortion and conjugate SI are attenuated to zero, unlike in traditional digital cancellation implementations. \emph{Note that when analyzing the proposed structure with actual waveform simulations in the later sections, digital SI cancellation is performed using a real least squares estimate of the coupling parameters, and thereby this assumption is not used there.}

\begin{table}[!t]
\renewcommand{\arraystretch}{1.3}
\caption{System level and general parameters of the proposed 2x2 MIMO full-duplex transceiver.}
\label{table:system_parameters}
\centering
\begin{tabular}{|c||c|c|c||c|}
\cline{1-2} \cline{4-5}
\textbf{Parameter} & Value & & \textbf{Parameter (cont.)} & Value\\
\cline{1-2} \cline{4-5}
SNR target & 10 dB & & RF cancellation & 30 dB\\
\cline{1-2} \cline{4-5}
Bandwidth & 12.5 MHz & & ADC bits & 12\\
\cline{1-2} \cline{4-5}
RX noise figure & 4.1 dB & & Ref. RX ADC bits & 12\\
\cline{1-2} \cline{4-5}
Sensitivity & -88.9 dBm & & PAPR & 10 dB\\
\cline{1-2} \cline{4-5}  
RX input power & -83.9 dBm & & IRR (TX) & 25 dB\\
\cline{1-2} \cline{4-5}
Antenna separation & 40 dB & & IRR (RX) & 60 dB\\
\cline{1-2} \cline{4-5} 
\end{tabular}
\end{table}

\begin{table}[!t]
\renewcommand{\arraystretch}{1.3}
\caption{Parameters for the relevant components of the transmitter, receiver, and reference receiver chains.}
\label{table:parameters}
\centering
\begin{tabular}{|c||c||c||c||c|}
\hline
\textbf{Component} & \textbf{Gain (dB)} & \textbf{IIP2 (dBm)} & \textbf{IIP3 (dBm)} & \textbf{NF (dB)}\\
\hline
PA (TX) & 27 & - & 15 & 5\\
\hline
LNA (RX) & 25 & - & 5 & 4.1\\
\hline
Mixer (RX)& 6 & 50 & 15 & 4\\
\hline
VGA (RX) & 0-69 & 50 & 20 & 4\\
\hline
\end{tabular}
\vspace{-4mm}
\end{table}

Furthermore, to obtain a more in-depth understanding regarding the performance of the proposed structure, power levels of the different signal components are compared in two scenarios: for the proposed reference receiver aided digital cancellation procedure, as well as for the traditional full-duplex transceiver structure utilizing linear digital cancellation. A block diagram for the latter structure can be found, e.g., in \cite{Korpi132}. The power levels for this scenario have been calculated based on the equations presented in \cite{Korpi13,Korpi133}, using the same parameters as presented in Tables~\ref{table:system_parameters} and~\ref{table:parameters}.

\subsubsection{Numerical Results}
\begin{figure}[!t]
\centering
\includegraphics[width=\columnwidth]{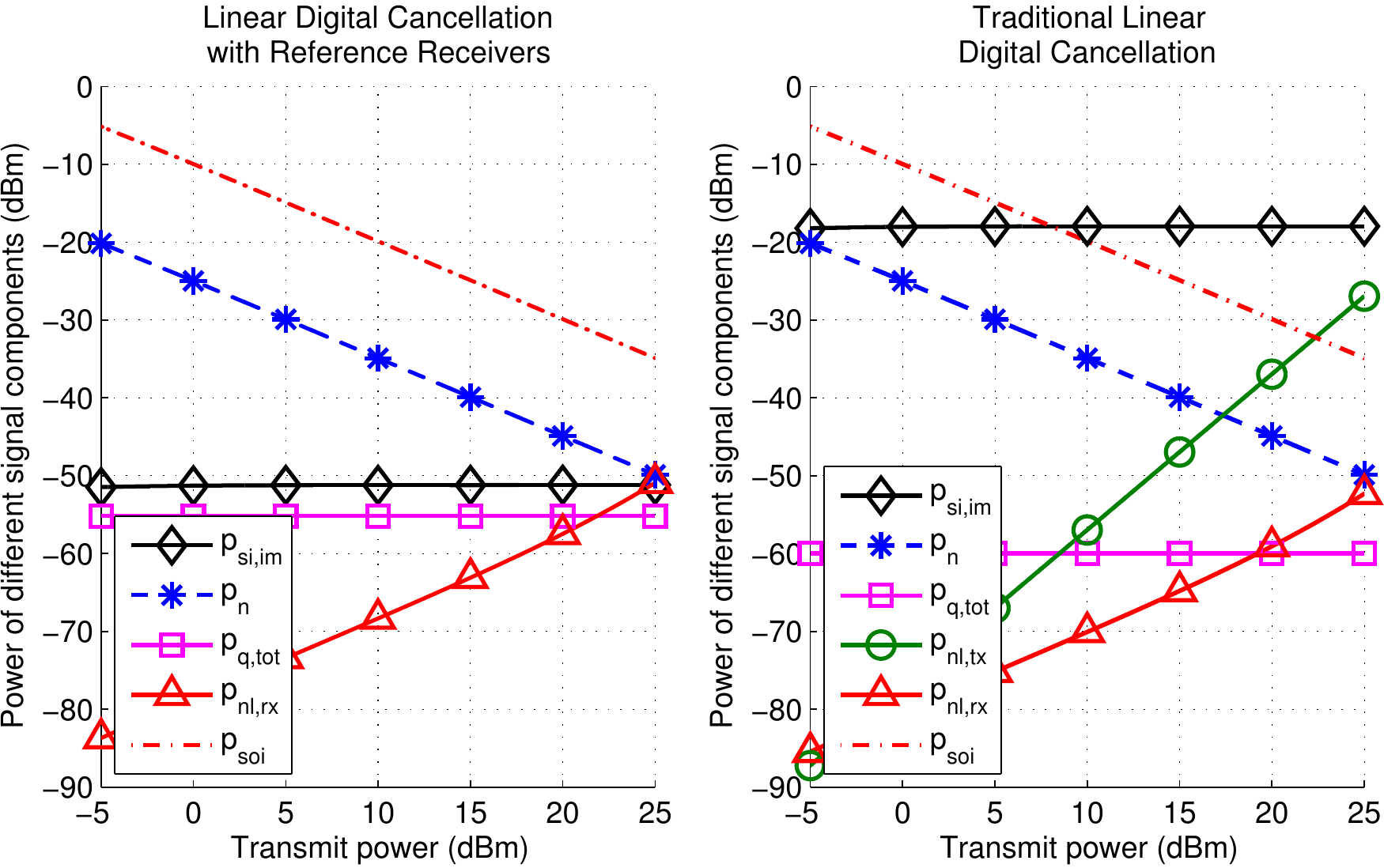}
\caption{The power levels of the different signal components at the detector input of an individual receiver chain.}
\label{fig:p_levels}
\vspace{-3mm}
\end{figure}

The resulting power levels, with respect to the transmit power of a single transmitter, are shown in Fig.~\ref{fig:p_levels}. Because of the AGC, the overall gain of each receiver chain is decreasing with increasing transmit powers to prevent the saturation of the ADC, and thus the absolute power levels of some signal components actually decrease when transmit power increases. When observing the power levels for the proposed reference receiver aided digital cancellation, it can be concluded that its signal-to-interference-plus-noise ratio (SINR) is mainly limited by the thermal noise floor ($p_n$). In fact, the RX-induced nonlinearities ($p_{nl,rx}$), alongside with IQ imaging ($p_{si,im}$), start affecting the SINR only with the highest transmit powers. This is obviously a promising preliminary result, since it indicates that the increased analog complexity of the proposed full-duplex transceiver structure indeed improves the achievable performance.

This finding is further emphasized by observing the power levels for the traditional linear digital cancellation, shown in the right-hand plot in Fig.~\ref{fig:p_levels}. It can be seen that, in this case, IQ imaging ($p_{si,im}$), alongside with TX-induced nonlinear distortion ($p_{nl,tx}$), deteriorates the SINR already with very low transmit powers. The reason for this lies in the reference signals for digital cancellation, as in this case they do not include any of the TX-induced distortion sources, unlike in the reference receiver aided digital cancellation. Another significant observation is that the level of the RX-induced nonlinear distortion is nearly the same for the reference receiver aided and traditional linear digital cancellation. This indicates that the reference RX chains do not produce intolerable amounts of nonlinear distortion to the reference signals. Overall, it can be concluded that the proposed reference receiver aided digital cancellation procedure performs well, despite deploying only linear cancellation processing. Furthermore, it will be shown below to outperform the existing linear and nonlinear solutions also in cases where practical cancellation parameter learning algorithms are deployed.

\subsection{Reference Receiver Aided Digital Cancellation Procedure}

As already discussed, the novelty of the proposed digital cancellation procedure lies in the reference signals, which are now taken from the outputs of the transmitters. This allows for the usage of linear digital processing, while still being able to cancel all the TX-induced distortion components of the SI signal. Below, we consider two different scenarios for digital SI cancellation, which are as follows.
\begin{enumerate}[I]
\item A separate calibration period for SI channel estimation during which there is no received signal of interest present.
\item No calibration period, which means that the SI channel estimation will be performed with the received signal of interest included in the total signal.
\end{enumerate}
Assuming uncorrelated transmit and receive signals, in the latter scenario the signal of interest will act as an additional noise source during the SI channel estimation.

To outline the actual channel estimation procedure, let us first determine the total received signal after the analog-to-digital conversion. It can be written for the $i$th RX chain as
\begin{align}
	y_{i,ADC}(n) = \sum_{j=1}^{N_{tx}} h_{ij}(n) \star x_{j,PA}(n) + z_i(n) \text{,} \label{eq:y_adc_n}
\end{align}
where $\star$ denotes convolution, $h_{ij}(n)$ is the impulse response of the effective linear channel experienced by the $j$th transmit signal $x_{j,PA}(n)$ when propagating to the $i$th receiver, and $z_i(n)$ represents the other sources of noise and distortion. Note that if there is no separate calibration period, $z_i(n)$ includes also the actual received signal of interest. To obtain the necessary channel estimates for digital SI cancellation, let us rewrite \eqref{eq:y_adc_n} with vector notation:
\begin{align}
	\mathbf{y}_{i,ADC} = \sum_{j=1}^{N_{tx}}  \mathbf{X}_{j,PA} \mathbf{h}_{ij} + \mathbf{z}_i \text{,} \label{eq:y_adc}
\end{align}
where $\mathbf{X}_{j,PA}$ is a covariance windowed convolution matrix of the form
\begin{align*}
\mathbf{X}_{j,PA} =  \left[\begin{smallmatrix}
  x_{j,PA}(M-1) & x_{j,PA}(M-2) & \cdots & x_{j,PA}(0) \\
  x_{j,PA}(M) & x_{j,PA}(M-1) & \cdots & x_{j,PA}(1) \\
  \vdots  & \vdots  & \ddots & \vdots  \\
  x_{j,PA}(N-1)& x_{j,PA}(N-2) & \cdots & x_{j,PA}(N-M)
 \end{smallmatrix}\right]
\end{align*}
and $\mathbf{h}_{ij}$ is of the form
\begin{align*}
	\mathbf{h}_{ij} = \begin{bmatrix} h_{ij}(0) & h_{ij}(1) & \cdots &  h_{ij}(M-1) \end{bmatrix}^{T} \text{,}
\end{align*}%
assuming a parameter estimation sample size of $N$ and a channel impulse response length of $M$. Using basic vector algebra, \eqref{eq:y_adc} can be further modified into
\begin{align}
	\mathbf{y}_{i,ADC} = \mathbf{X}_{PA} \mathbf{h}_i + \mathbf{z}_i \text{,} \label{eq:y_adc_simple}
\end{align}
where $\mathbf{X}_{PA} = \begin{bmatrix} \mathbf{X}_{1,PA} & \mathbf{X}_{2,PA} & \cdots &  \mathbf{X}_{N_{tx},PA} \end{bmatrix}$, and $\mathbf{h}_i = \begin{bmatrix} \mathbf{h}_{i1}^T & \mathbf{h}_{i2}^T & \cdots &  \mathbf{h}_{iN_{tx}}^T \end{bmatrix}^T$.

%\begin{align*}
%\mathbf{X}_{PA} &= \begin{bmatrix} \mathbf{X}_{1,PA} & \mathbf{X}_{2,PA} & \cdots &  \mathbf{X}_{N_{tx},PA} \end{bmatrix} \text{, and}\nonumber\\
%\mathbf{h}_i &= \begin{bmatrix} \mathbf{h}_{i1}^T & \mathbf{h}_{i2}^T & \cdots &  \mathbf{h}_{iN_{tx}}^T \end{bmatrix}^T \text{.}
%\end{align*}

To construct then the necessary equations for channel estimation, it still remains to determine the matrix $\mathbf{X}_{PA}$. This can be done simply by substituting $x_{j,PA}(n)$ with $x_{j,ref}(n)$, which denotes here the $j$th reference receiver observation. Thus, by setting $x_{j,PA}(n) = x_{j,ref}(n)$, we can construct an approximation of the total convolution matrix $\mathbf{X}_{PA}$, which we will denote by $\mathbf{X}_{ref}$. Then, as the $i$th received signal $\mathbf{y}_{i,ADC}$ is obviously known, it is trivial to obtain estimates for the channel responses based on \eqref{eq:y_adc_simple}. In this study, we choose to use least squares for the actual estimation, as it is a simple and robust method for this kind of a problem. The channel estimates are thus given by
\begin{align}
	\mathbf{\hat{h}}_i = (\mathbf{{X}}_{ref}^H \mathbf{{X}}_{ref} )^{-1}\mathbf{{X}}_{ref}^H \mathbf{y}_{i,ADC} \text{,} \label{eq:ch_est}
\end{align}
where $()^H$ denotes the Hermitian transpose.

Utilizing \eqref{eq:ch_est}, the signal after the actual digital SI cancellation can then be written as follows.
\begin{align*}
	y_{i,DC}(n) = y_{i,ADC}(n)-\sum_{j=1}^{N_{tx}} \hat{h}_{ij}(n) \star x_{j,ref}(n) + z_i(n) \text{,}
\end{align*}
where each $\hat{h}_{ij}(n)$ is obtained from $\mathbf{\hat{h}}_i$. It should also be noted that the outlined estimation procedure is the same with and without a special calibration period. Furthermore, this same procedure can also be used for traditional digital SI cancellation implementations, but then the signals $x_{j,ref}(n)$ consist of the original transmitted data samples.

\subsubsection*{Lower Bound on Channel Estimator Variance}
% Painota CRLB:n yleisyyttä?
To derive an approximation for the lower bound of the SI channel estimator variance, let us assume that the powers of RX-induced nonlinear distortion and SI mirror image are negligibly low. This is of course not exactly true with higher transmit powers but making this assumption allows us to determine the Cram\'{e}r--Rao lower bound (CRLB) for the estimator. This, on the other hand, reveals the effect of the signal of interest on the quality of the channel estimate when there is no separate calibration period. {\color{black}Let us further assume that OFDM signals are used for both transmission and reception.} Now, based on the approximation that OFDM signals are normally distributed with a sufficiently large number of subcarriers \cite{Shuangqing10}, the distribution of $\mathbf{z}_i$ is approximately multivariate Gaussian. Thus, without a separate calibration period, the CRLB for any SI channel estimate can be derived as \cite{Kay93}
\begin{align}
	\operatorname{Cov}(\mathbf{\hat{h}}_i) &\geq (\mathbf{{X}}_{ref}^H \mathbf{{X}}_{ref})^{-1} (p_{soi}+p_{n})\nonumber\\
	 &\approx \left(\frac{p_{soi}+p_{n}}{N}\right) \mathbf{R}_{x,ref}^{-1} \label{eq:var_h_1}\text{,}
\end{align}
where the latter expression is based on the notion that, with a large parameter estimation sample size $N$, $\mathbf{{X}}_{ref}^H \mathbf{{X}}_{ref}$ can be approximated with ensemble augmented covariance matrix $\mathbf{R}_{x,ref}$ of reference receiver observation data $x_{j,ref}(n)$, that is, $\mathbf{{X}}_{ref}^H \mathbf{{X}}_{ref} \approx N \mathbf{R}_{x,ref}$ \cite{Schreier10}. Furthermore, assuming that consecutive samples of the transmit signals are uncorrelated, and that each transmitter is using the same transmit power, it can be written that $\mathbf{R}_{x,ref} \approx p_{ref} \mathbf{I}$, where $\mathbf{I}$ is an identity matrix of the same dimensions as $\mathbf{R}_{x,ref}$, and $p_{ref}$ is a constant. Substituting this into \eqref{eq:var_h_1}, we get
\begin{align}
	\operatorname{Cov}(\mathbf{\hat{h}}_i) &\geq \left(\frac{p_{soi}+p_{n}}{N p_{ref}}\right) \mathbf{I} \label{eq:var_h_2}\text{,}
\end{align}
or for each individual tap of the channel estimate:
\begin{align}
	\operatorname{Var}(\hat{h}_{ij}) &\geq \left(\frac{p_{soi}+p_{n}}{N p_{ref}}\right) \label{eq:var_h_ij}\text{.}
\end{align}

When considering a situation under a separate calibration period, we can write $p_{soi} = 0$. Thus, in this case the CRLB becomes
\begin{align}
	\operatorname{Var}(\hat{h}_{ij,c}) &\geq \left(\frac{p_{n}}{N_c p_{ref}}\right) \label{eq:var_h_ijc}\text{,}
\end{align}
where the subscript $c$ indicates a case under a separate calibration period. Using \eqref{eq:var_h_ij} and \eqref{eq:var_h_ijc}, it is possible to determine how the presence of the signal of interest affects the parameter estimation sample size required to calculate the SI channel estimate. Namely, by requiring a similar variance for each tap as achieved under a separate calibration period using $N_c$ parameter estimation samples, i.e., $\operatorname{Var}(\hat{h}_{ij}) = \operatorname{Var}(\hat{h}_{ij,c})$, the needed number of parameter estimation samples without a calibration period can easily be shown to be
\begin{align}
	N &= N_c \left(\mathit{snr}+1 \right) \text{,} \label{eq:N_req}
\end{align}
where $\mathit{snr} = \frac{p_{soi}}{p_{n}}$ is the signal-to-noise ratio at the detector input. Thus, assuming an example SNR of, say, 15 dB, approximately 33 times more parameter estimation samples are required to achieve a similar accuracy for the SI channel estimate if the received signal of interest is also present in the total signal. Note that this should not be interpreted as an exact relation between the required parameter estimation sample sizes in the two scenarios, as the CRLB can rarely be achieved under typical circumstances. Nevertheless, \eqref{eq:N_req} still provides useful insight into how the signal of interest affects the accuracy of the SI channel estimate, which has not been analyzed earlier in the full-duplex literature.

%From \eqref{eq:var_h_ij} it can be observed that the signal of interest will increase the variance of the channel estimate, as expected. Namely, with a separate calibration period, we could write that $p_{soi} = 0$, which would obviously decrease the variance. However, 

\vspace{-1mm}
\section{Full Waveform Simulations}
\label{sec:simulations}

To assess the actual realized performance of the proposed structure for a MIMO full-duplex transceiver, waveform simulations are performed. The model presented in Fig.~\ref{fig:block_diagram} is assumed in the simulations, with OFDM waveforms for the signals. The used waveform simulator models each component explicitly using baseband equivalent models, which include the effects of all the considered circuit impairments without any approximations. These baseband equivalent models are constructed according to the parameters presented in Tables~\ref{table:system_parameters} and~\ref{table:parameters}, with additional parameters for the waveforms being shown in Table~\ref{table:simul_param}. The actual channel estimate for digital SI cancellation is calculated with least squares, using \eqref{eq:ch_est}. Thus, we make no assumptions regarding the accuracy of the channel estimate, but instead calculate it under realistic conditions. 

In this analysis, \emph{the SINR after digital SI cancellation is used as the performance metric}. The SINR is evaluated and analyzed for the proposed structure utilizing reference RX chains, as well as for the traditional method where the reference signals consist of the original transmitted samples. In addition, the SINRs achieved with widely-linear and nonlinear digital cancellation, described in \cite{Korpi133} and \cite{Anttila13}, are also shown for comparison, the corresponding SI channel estimates being again calculated with least squares.

%\begin{table}[!t]
%\renewcommand{\arraystretch}{1.3}
%\caption{Additional waveform simulator parameters.}
%\label{table:simul_param}
%\centering
%\begin{tabular}{|c||c|}
%\hline
%\textbf{Parameter} & \textbf{Value}\\
%\hline
%Constellation & 16-QAM\\
%\hline
%Number of subcarriers & 64\\
%\hline
%Number of data subcarriers & 48\\
%\hline
%Guard interval & 16 samples\\
%\hline
%Sample length & 15.625 ns\\
%\hline
%Symbol length & 4 $\mu$s\\
%\hline
%Signal bandwidth & 12.5 MHz\\
%\hline
%Oversampling factor & 4\\
%\hline
%K-factor of the SI channel & 35.8 dB\\
%\hline
%\end{tabular}
%\vspace{-1mm}
%\end{table}

\begin{table}[!t]
\renewcommand{\arraystretch}{1.3}
\caption{Additional waveform simulator parameters.}
\label{table:simul_param}
\centering
\begin{tabular}{|c||c|c|c||c|}
\cline{1-2} \cline{4-5}
\textbf{Parameter} & Value & & \textbf{Parameter (cont.)} & Value\\
\cline{1-2} \cline{4-5}
Signal bandwidth & 12.5 MHz & & Symbol length & 4 $\mu$s\\
\cline{1-2} \cline{4-5}
Subcarriers & 64 & & Guard interval & 16 samples\\
\cline{1-2} \cline{4-5}
Data subcarriers & 48 &  & Constellation & 16-QAM\\
\cline{1-2} \cline{4-5}
Oversampling factor & 4 & & K-factor of the & 35.8 dB\\
\cline{1-2}
Sample length & 15.625 ns & & SI channel &  \\
\cline{1-2} \cline{4-5}
\end{tabular}
\vspace{-5mm}
\end{table}

\begin{figure}[!t]
\centering
\includegraphics[width=\columnwidth]{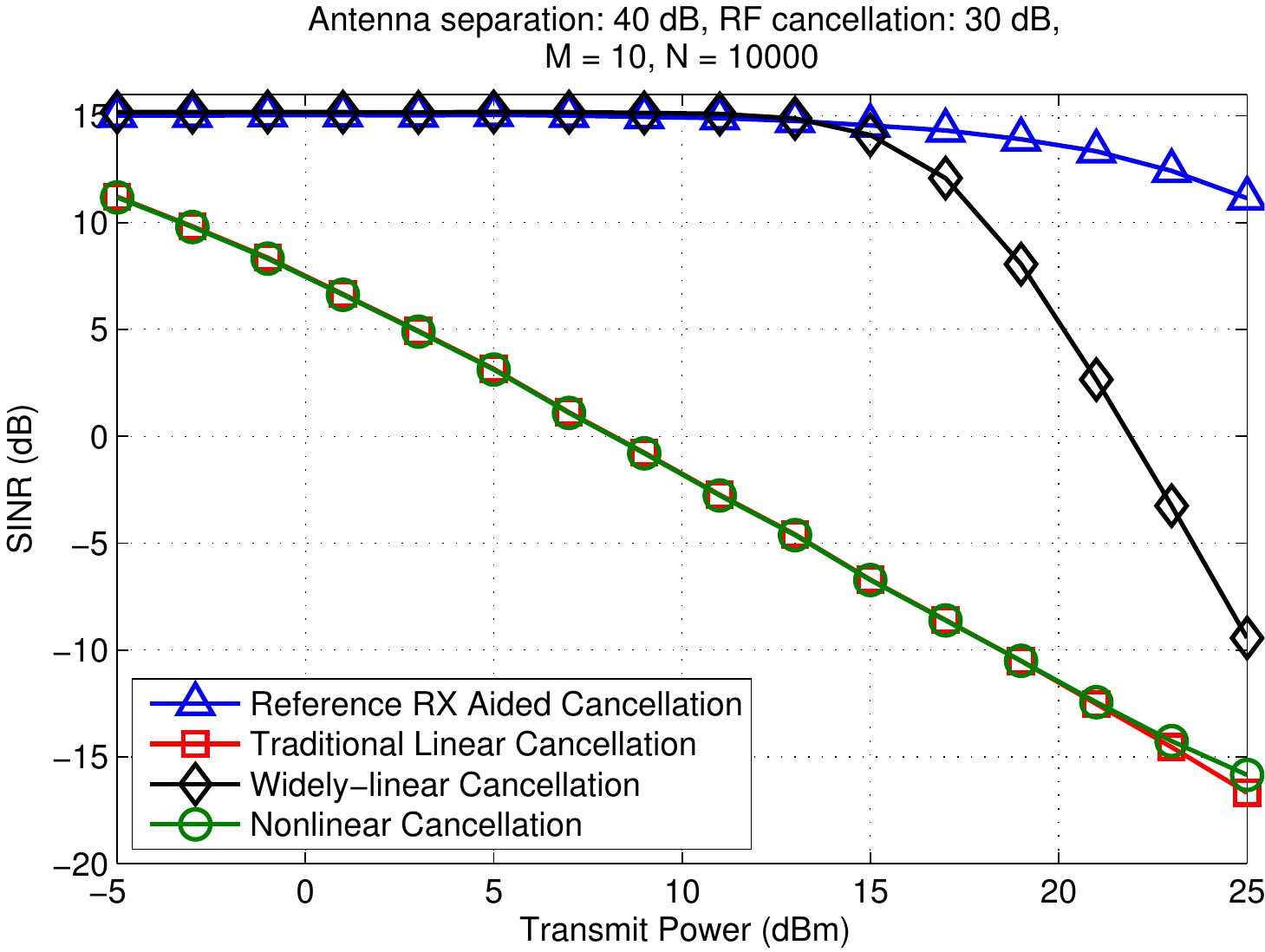}
\caption{The simulated SINRs with the proposed structure, traditional linear digital cancellation, widely-linear digital cancellation, and nonlinear digital cancellation.}
\label{fig:sinr}
\vspace{-5mm}
\end{figure}

The resulting SINRs at the detector input of an individual RX chain, with respect to the transmit power of a single transmit antenna, are shown in Fig.~\ref{fig:sinr}. This figure has been generated assuming a parameter estimation sample size of 10000 samples and a separate calibration period. It can be observed that the traditional method, where the transmitted samples are linearly processed to produce the digital cancellation signal, achieves by far the lowest SINR. Similarly, the SINR achieved with nonlinear digital cancellation is almost equally low. This is mostly due to the conjugate SI, caused by the IQ imbalance occurring in the TX chain, which is a dominating source of distortion in this scenario. Thus, it is clear that neither linear nor nonlinear digital cancellation is sufficient to cope with the residual SI in a realistic MIMO full-duplex transceiver.

For the proposed reference RX chain aided digital cancellation and widely-linear digital cancellation, TX-induced conjugate SI is not a bottleneck. For the former, this conjugate SI is automatically included in the reference signals for digital cancellation, and it is thus attenuated also in the digital domain. In the latter scenario, the conjugate SI is explicitly modelled in the digital domain, as it is not included in the reference signals as such. In addition, widely-linear digital cancellation has the benefit that it is able to mitigate also the conjugate SI produced by the RX chain, although in this analysis we assume that the receiver IQ mixer is well calibrated and has a fairly high image rejection ratio. However, these benefits come with the cost of higher computational complexity, as the response for the conjugate SI must also be estimated, unlike in the proposed reference RX chain -based structure \cite{Korpi133}.

These deductions are confirmed by Fig.~\ref{fig:sinr}, where it can be observed that both the widely-linear and reference receiver -based implementations achieve nearly the ideal SINR of 15~dB with transmit powers below 15~dBm. This is due to the ability to attenuate the TX-induced conjugate SI. However, with higher transmit powers, the SINR decreases for both of these cancellation solutions. This is caused by nonlinear distortion, which becomes considerably powerful with higher transmit powers. The decrease in the SINR is steeper for the widely-linear digital cancellation, as it is not able to model the nonlinearities in any way. The reference RX chain based digital canceller, on the other hand, is able to attenuate the TX-incuded nonlinear distortion, as it is included in the reference signals. Thus, with very high transmit powers, it achieves the highest SINR of all the considered techniques, its performance being limited only by the nonlinearities and conjugate SI produced in the actual and reference RX chains.

To observe the effect of the parameter estimation sample size ($N$) on the proposed reference receiver aided digital cancellation scheme, next we will fix the transmit power to 15~dBm and vary $N$. This provides information regarding the required number of parameter estimation samples for accurate SI channel estimation. In addition, the SINR is calculated also for a situation where there is no separate calibration period. This means that the SI channel estimation will be done while the actual signal of interest is being received.

\begin{figure}[!t]
\centering
\includegraphics[width=\columnwidth]{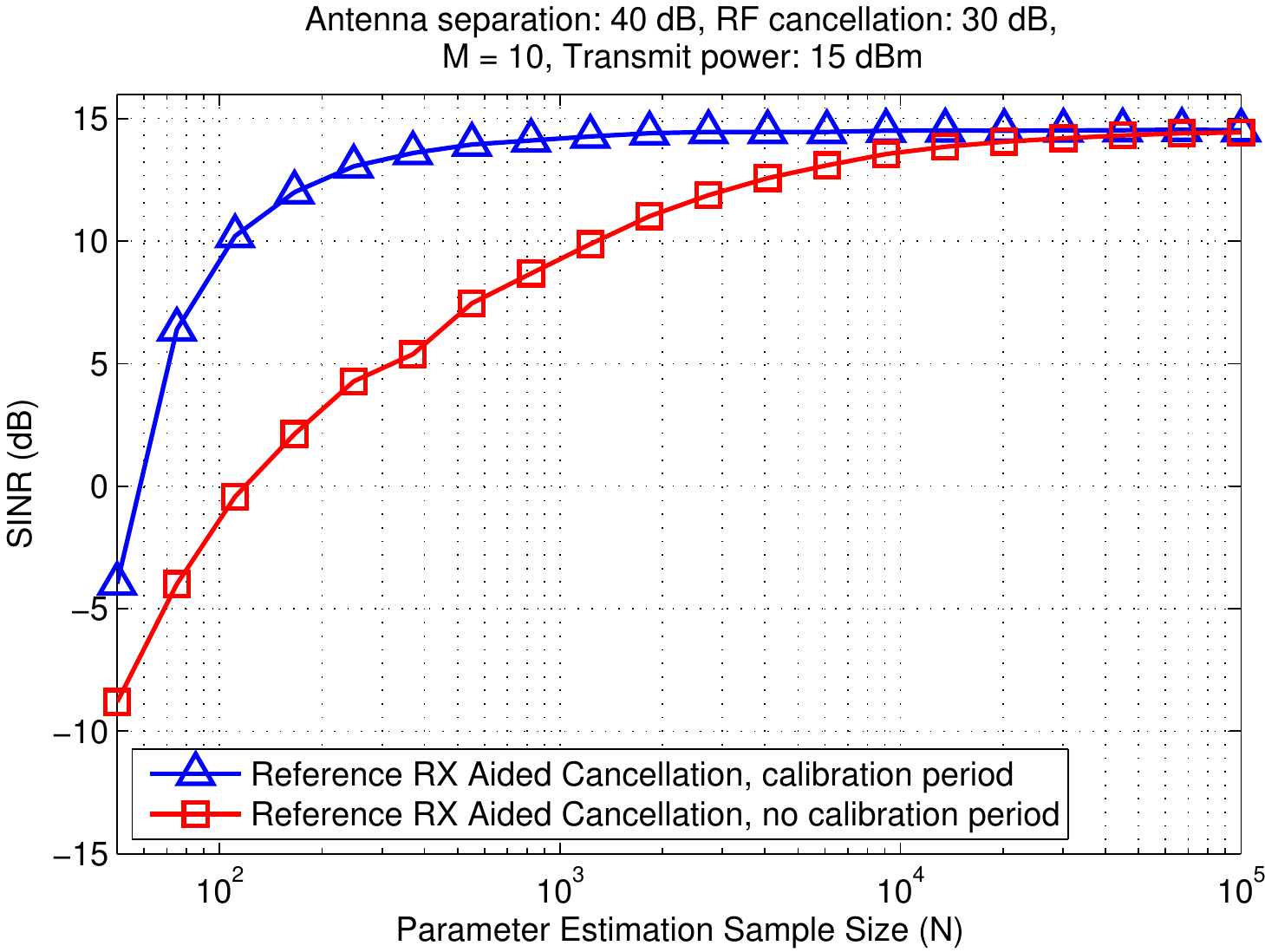}
\caption{The simulated SINRs with respect to the parameter estimation sample size, determined with and without a calibration period.}
\label{fig:sinr_n}
\vspace{-4mm}
\end{figure}

The corresponding SINRs are shown in Fig.~\ref{fig:sinr_n}. It can be observed that, with a separate calibration period, it is possible to achieve the highest possible SINR with approximately 4000 parameter estimation samples. Note that with this transmit power it is not possible to exactly achieve the ideal SINR of 15 dB as the RX-induced distortion components are already increasing the noise floor.

With the chosen parameters, \eqref{eq:N_req} predicts that approximately 130000 parameter estimation samples are required to achieve similar performance without a calibration period. When observing Fig.~\ref{fig:sinr_n}, it can be seen that, with a parameter estimation sample size of 100000, the SINR corresponding to no calibration period is almost as high as the SINR achieved with a separate calibration period. This indicates that slightly more than 100000 parameter estimation samples are required for the SINRs to be equal, which matches rather well with the prediction given by \eqref{eq:N_req} and confirms the validity of the assumptions made when deriving it. This is also an encouraging result in itself, as it is highly beneficial to be able to do the SI channel estimation without the additional overhead presented by a calibration period. Thus, Equation~\eqref{eq:N_req} and Fig.~\ref{fig:sinr_n} suggest that, by using a sufficient parameter estimation sample size, it is indeed possible to perform digital SI cancellation without such overhead.

\vspace{-1mm}
\section{Conclusion}
\label{sec:conc}

In this paper we have proposed and analyzed a novel structure for an in-band MIMO full-duplex transceiver, which utilizes additional receivers to obtain the reference signals for digital cancellation from the outputs of the transmitter chains. The benefit of this approach lies in the fact that now the digital cancellation signal automatically includes all the sources of transmitter-induced distortion, and thus they will be attenuated by digital cancellation even when utilizing only linear digital processing. The proposed reference receiver based digital cancellation scheme was evaluated with full waveform simulations, where it was shown that it achieves higher SINRs than the existing linear and nonlinear digital cancellation solutions throughout the considered transmit power range. The drawback of this scheme, however, is its increased analog complexity, {\color{black}and more work is required to determine whether the proposed method is actually the most efficient way of exploiting the additional receiver chains}. Nevertheless, if only linear digital processing can be utilized, the proposed reference receiver aided digital cancellation scheme is a viable option to increase the amount of self-interference attenuation.

%\vspace{-3mm}
\bibliographystyle{./IEEEtran}
\bibliography{./IEEEabrv,./IEEEref}

\end{document}